\documentclass[12pt]{iopart}

\newcommand{\profound}{{\sc ProFound}}
\newcommand{\propane}{{\sc Propane}}
\newcommand{\protools}{{\sc ProTools}}
\newcommand{\R}{{\sc R}}
\newcommand{\Rfits}{{\sc Rfits}}
\newcommand{\Rwcs}{{\sc Rwcs}}

\usepackage[numbers]{natbib}
\usepackage{graphicx}
\usepackage{color}
\usepackage{alltt}
\usepackage{enumitem}
\usepackage{amstext}

\setlist[enumerate,1]{label={(\arabic*)}}
\setlist[enumerate,2]{label={(\roman*)}}

\begin{document}

\title[JWST NIRCam Wisp Removal]{Dynamic Wisp Removal in JWST NIRCam Images}

\author[A.~S.~G. Robotham et al.]{
	A.~S.~G. Robotham,$^{1,2}$
	J.~C.~J. D'Silva,$^{1,2}$
	R.~A. Windhorst,$^3$ \\
	R.~A. Jansen,$^3$
	J. Summers,$^3$
	S.~P. Driver,$^1$ \\
	C.~N.~A. Wilmer,$^4$
	S. Bellstedt,$^{1,2}$
	}

\address{
	$^{1}$ICRAR, M468, University of Western Australia, Crawley, WA 6009, Australia \\
	$^{2}$ARC Centre of Excellence for Astrophysics in Three Dimensions (ASTRO3D) \\
	$^{3}$School of Earth and Space Exploration, Arizona State University, Tempe, AZ 85287-1404, USA\\
	$^{4}$Steward Observatory, University of Arizona, 933 N. Cherry Avenue, Tucson, AZ 85721-0009, USA
}

Exact ordering to be determined (above I am adding people as they respond).

\ead{aaron.robotham@uwa.edu.au}
\vspace{10pt}
\begin{indented}
\item[]April 2023
\end{indented}

\begin{abstract}
The James Webb Space Telescope (JWST) near-infrared camera (NIRCam) has been found to exhibit serious wisp-like structures in four of its eight short-wavelength detectors. The exact structure and strength of these wisps is highly variable with the position and orientation of JWST, so the use of static templates is non-optimal. Here we investigate a dynamic strategy to mitigate these wisps using long-wavelength reference images. Based on a suite of experiments where we embed a worst-case scenario median-stacked wisp into wisp-free images, we define suitable parameters for our wisp removal strategy. Using this setup we re-process wisp-affected public Prime Extragalactic Areas for Reionization and Lensing Science (PEARLS) data in the North Ecliptic Pole Time Domain Field (NEP-TDF), resulting in significant visual improvement in our detector frames and reduced noise in the final stacked images.
\end{abstract}

%
%
%
%
%

\section{Introduction}

\begin{figure}[]
\begin{center}
\includegraphics[width=6in]{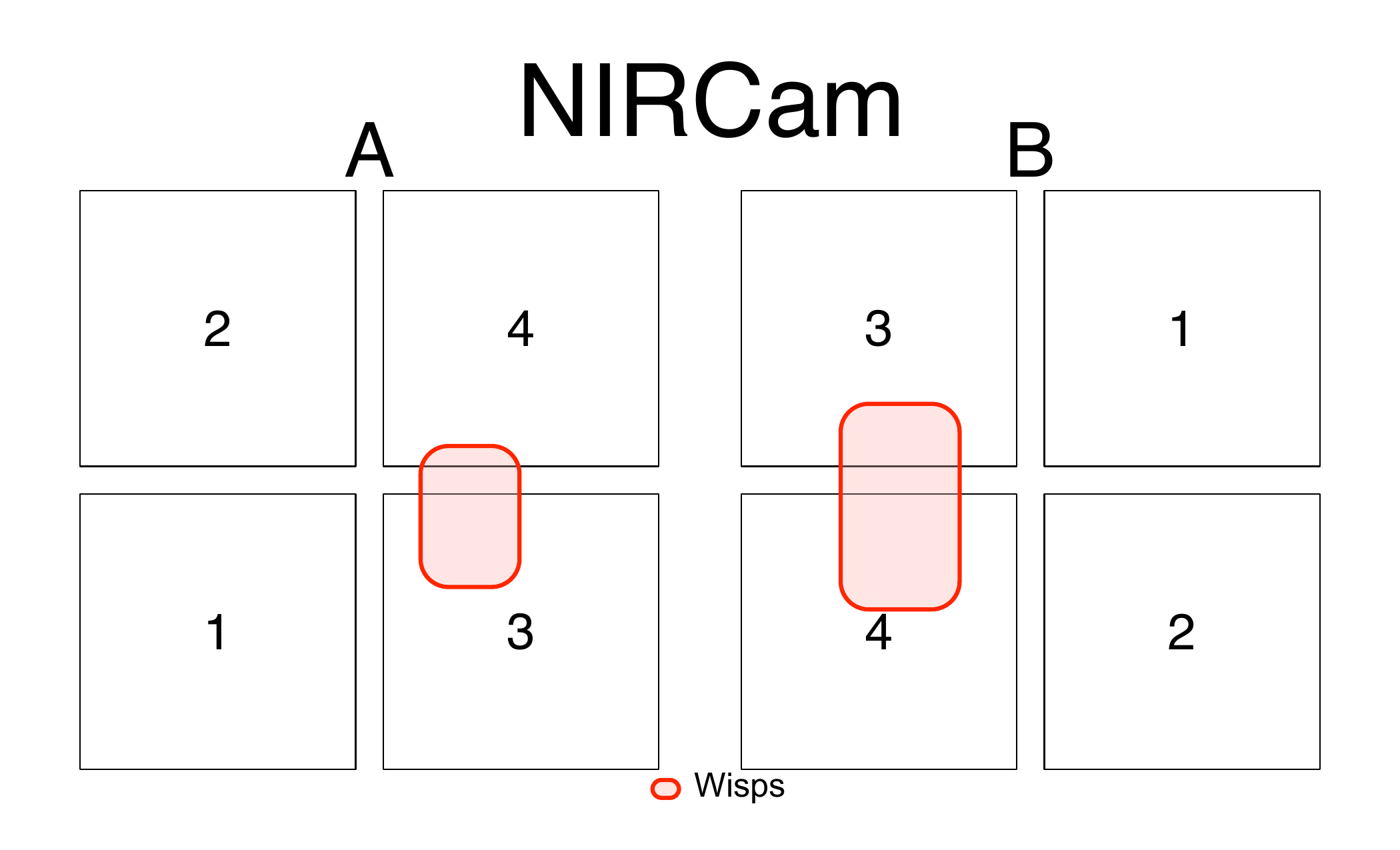}
\caption{Schematic layout of NIRCam short-wavelength modules (A LHS; B RHS) and their individually numbered detectors (1--4). {The most common way to refer to detectors is NRC[A--B][1--4] etc (i.e.\ `NRC' followed by the module letter, then detector number)}. The intra- and inter-module gaps are not to exact scale. Wisps are most visually prevalent in detectors NRCA3, NRCA4 and NRCB3 and NRCB4 (i.e.\ the central detectors), as indicated by the red regions.}
\label{fig:NIRCam_short}
\end{center}
\end{figure}

The James Webb Space Telescope (JWST) is a truly exceptional near- and mid-infrared facility \cite{2005SPIE.5904....1R}, with multiple major surveys being conducted and proposed using its wide-field near-infrared instrument NIRCam \cite{2023AJ....165...13W, 2023ApJ...946L..13F, 2022ApJ...935..110T}. Not unusually for a new facility, various facets of the data are slowly being understood and their calibration and mitigation improved by the astronomical community, e.g.\ cosmic rays, image persistence, dragon's breath, claws and snowballs etc \cite{2023PASP..135b8001R}\footnote{https://jwst-docs.STScI.edu/jwst-near-infrared-camera/nircam-features-and-caveats}. One of the most difficult to detect and remove are `wisps'. Wisps are present as cloud-like cirrus features on certain combinations of NIRCam filters and detectors. For reference, Figure \ref{fig:NIRCam_short} shows the approximate layout of the NIRCam short-wavelength modules and detectors (where wisps are prevalent).

Wisps appear to have a scattered-light origin, predominantly due to off-axis light from bright stars hitting the top secondary mirror strut and entering NIRCam through the aft-optic system. {Because the short-wavelength and long-wavelength cameras are on very different light paths, wisps can only appear in short-wavelength NIRCam data.} The STScI web pages suggest that the wisps should only significantly vary in strength for a given filter-detector combination\footnote{https://jwst-docs.STScI.edu/jwst-near-infrared-camera/nircam-features-and-caveats/nircam-claws-and-wisps}, however experience with Prime Extragalactic Areas for Reionization and Lensing Science \cite[PEARLS; ][]{2023AJ....165...13W} data reveals that wisp intensity and geometry can change quite significantly even for small (hundreds of pixel) dither offsets of a given filter-detector combination. This makes constructing static templates a potential dead-end for robust frame level removal as they are strongly sensitive to the exact geometry, intensity, and spectrum of off-axis light.

\begin{figure}[]
\begin{center}
\includegraphics[width=3in]{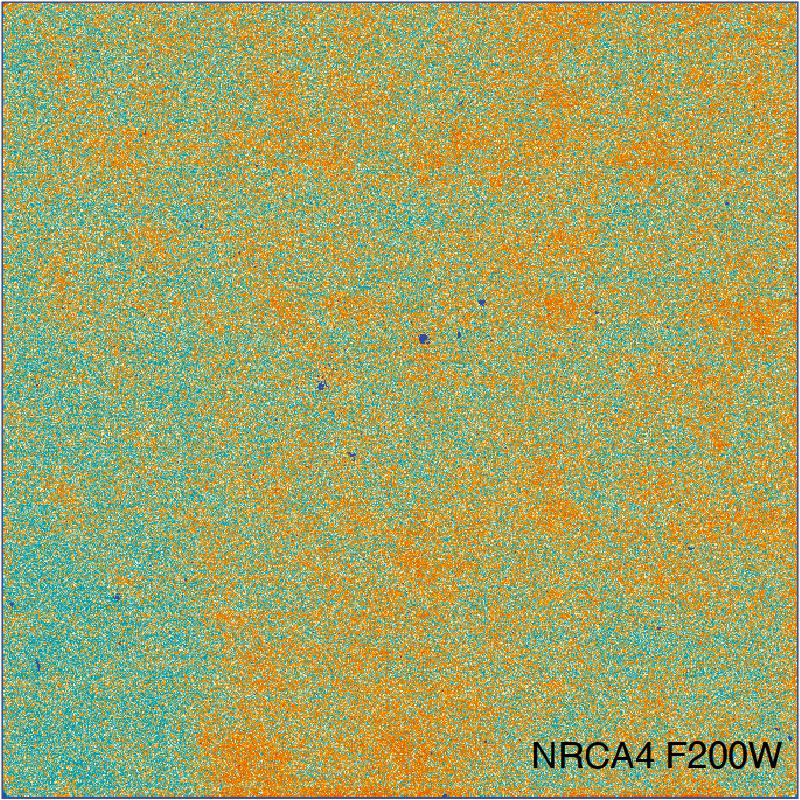}
\includegraphics[width=3in]{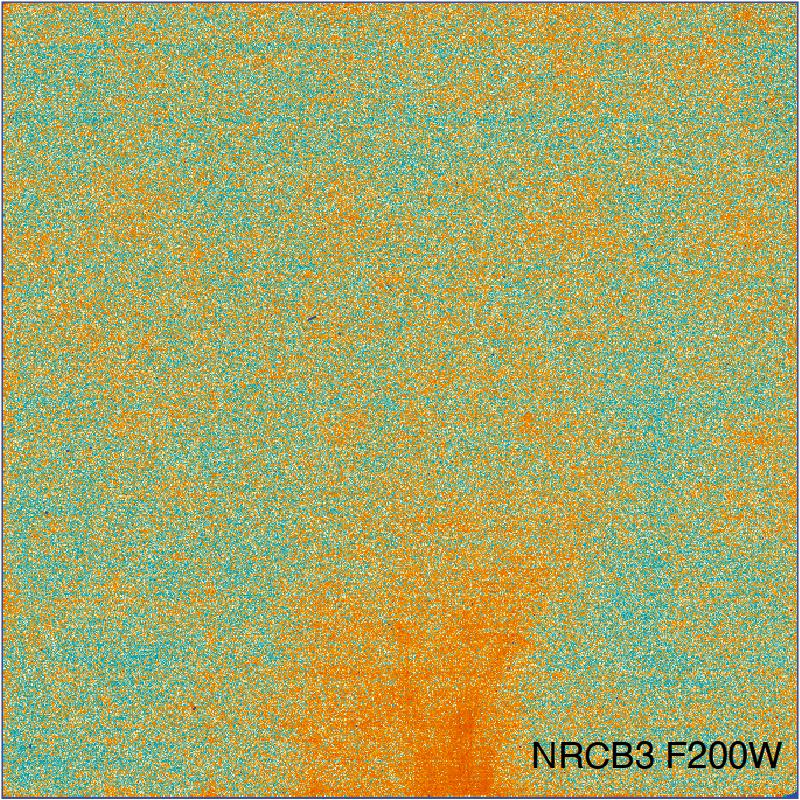} \\
\includegraphics[width=3in]{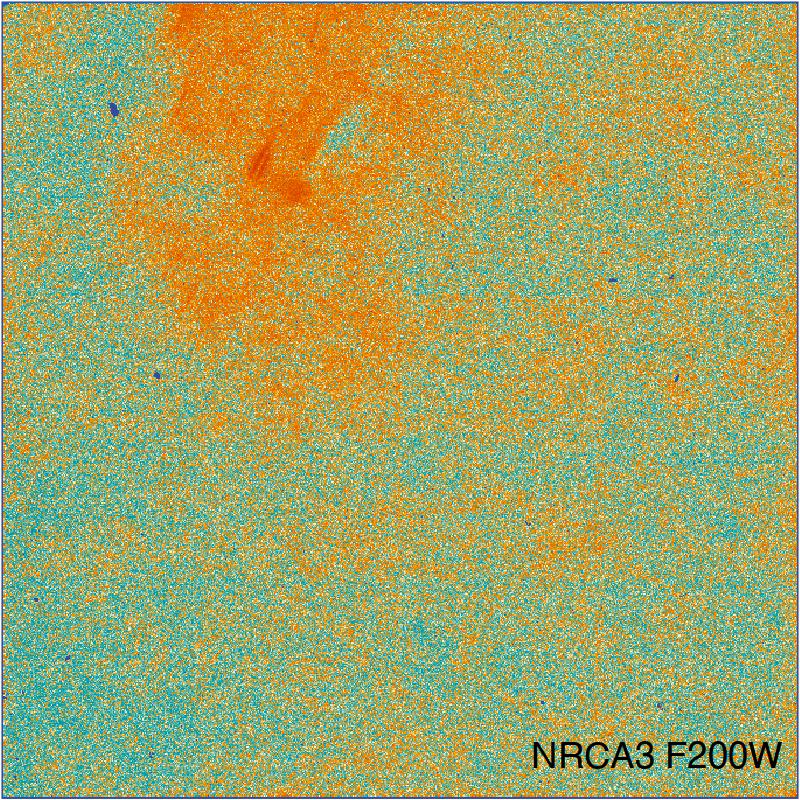}
\includegraphics[width=3in]{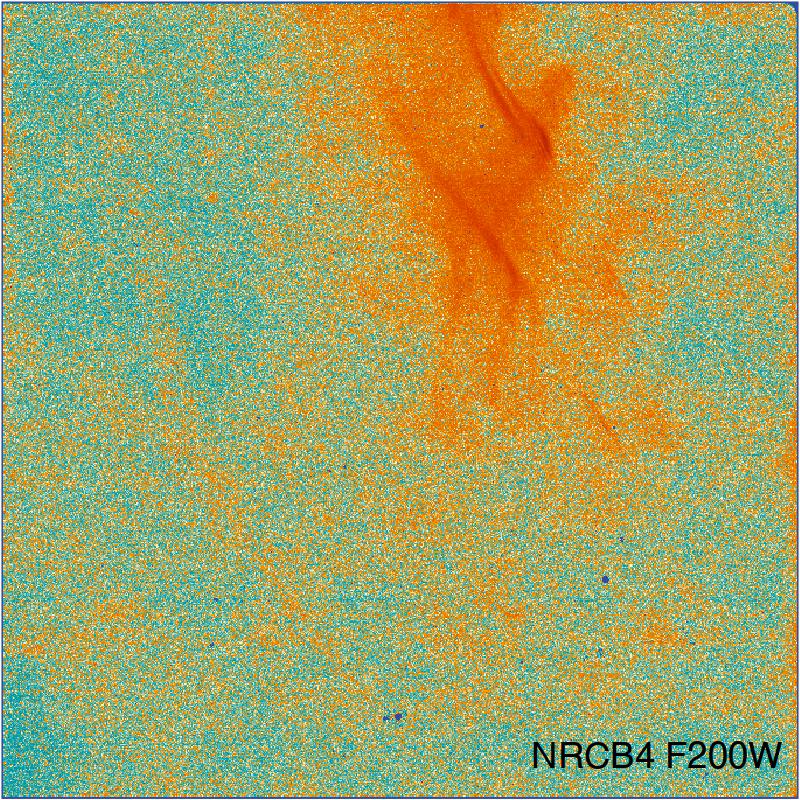}
\caption{STScI wisp templates created for {NRCA4} (top-left); {NRCA3} (bottom-left); {NRCB3} (top-right) and ({NRCB3} (bottom-right). The layout is chosen because it approximately reflects the true detector layout geometry of NIRCam (see the middle frames of Figure \ref{fig:NIRCam_short}, and wisp regions highlighted). Each image is median-subtracted and displayed with a blue-yellow-red asinh ramp (blue showing below-median pixels and red above-median pixels), with the wisps being the prominent central red features crossing {NRCA3--4} and {NRCB3--4}. Whilst {NRCB4} shows the most dramatic wisp structure, even in {NRCA4} the wisp amplitude is sufficient to create artefacts in large stacks of images with small dithers.}
\label{fig:STScI_wisp_templates}
\end{center}
\end{figure}

{As well as wisps only being present in short-wavelength NIRCam filter images (i.e.\ they are not present in F250M and long-wards), they only appear in certain detector locations (see Figure \ref{fig:STScI_wisp_templates}; wisps are apparent in the middle four detectors)}. NRCB4 has by far the most pronounced wisps when present, where the embedded filamentary features can sometimes achieve galaxy-like surface brightnesses. In general the F200W filter has the most serious artefacts from wisps, followed by F150W and often (to a much lesser extent) F115W and F090W {(in PEARLS we have less direct experience with the remaining short-wavelength filters, but strong wisps are known to appear in F182M)}. The fact that the geometry of the wisp varies so much with wavelength explains why the wisps vary even in a given filter. Clearly the spectrum of the off-axis nuisance source changes the integrated wisp pattern even within a filter. We note that STScI has only created wisp templates for filters F150W, F150W2, F200W and F210M (created by CW using commissioning plus public data up to late August 2022), so {F070W, F090W, F115W, F140M, F162M, F182M (the remaining short-wavelength channel medium and wide filters)} in particular need a remedial strategy despite cases where the other filter templates appear to work well at partly mitigating the wisps. Even where STScI templates are available, there remain quite a few artefacts (hot and cold pixels) and low signal-to-noise data in much of the frame.{Further to this, using the templates still requires scaling to match the image wisps present, so some care regarding source masking and template subtraction is necessary.}

Given the above statements it is intuitively clear that some programs and visits will have fewer wisp issues (due to the good fortune of exactly where off-axis sources happen to lie), and others might be seriously compromised. In the PEARLS medium-deep survey fields there are examples of detector-filter combinations with no visual wisps apparent in our final stacks, and other cases where wisps are highly prominent and problematic for further analysis. This paper presents a dynamic strategy for optimally removing wisps. We believe this approach will prove useful to the broader community for future medium and deep surveys with NIRCam and help to achieve $\sqrt{T}$ improvements in depth with increasing exposure time ($T$) across all NIRCam filters.

\section{Wisp Removal}

Having identified wisp residuals in inverse-variance-weighted and median stacks of PEARLS fields, it became clear that better dynamic strategies were necessary for us to extract maximum science from the observations. This was particularly true for the PEARLS Northern Ecliptic Pole Time Domain Field (NEP-TDF), which due to unfortunate coincidence is particularly affected by wisps. Given it will be one of the deeper wide JWST fields, it is paramount we mitigate their presence as much as possible.

First efforts in this direction relied on the automated sky detection capabilities of \profound{} as detailed in \cite{2018MNRAS.476.3137R, 2022AJ....164..141W} and \cite{2023AJ....165...13W}. This worked well for some filter-detector-field combinations, where the wisp happened to have quite smooth and extended characteristics. The best metric of quality in these scenarios is the Normality of the nominal `sky' background pixels, since a well-characterised sky (both in terms of background and root-mean-squared, RMS, deviation) should have a mean of zero and a standard deviation of one. It was clear that some filter-detector-field combinations had features not captured by the smooth sky modelling assumptions that underly \profound{}'s background subtraction methods. Most seriously, bright streaks in {NRCB4} have far too high a spatial frequency to be modelled as sky without creating serious self-subtraction issues.

\subsection{Colour Difference Flagging}
\label{sec:process}

When analysing large quantities of stacked and mosaicked NIRCam data, a consistent finding was that wisps, whilst present to varying degrees in all of the short-wavelength detector combinations, are \emph{never} present in the long-wavelength detectors (F250M and long-wards). {Wisp features, and their behaviour with wavelength, was also noted in} \cite{2023PASP..135d8001R}. This offers a conceptual solution to identifying wisps in a dynamic manner: if features are only present in a short-wavelength filter frame compared to a reference long-filter stacked image of similar or greater depth, then it is likely the feature is a scattered light wisp rather than an extremely blue object. {In fact, very few real objects should exist with colours in JWST filters bluer than an OB star — a moderate redshift ($0.8 < z < 2$) H$\alpha$ bright source being one of the exceptional cases \cite[e.g.\ see][]{2012MNRAS.420..878K}.}

After some experimentation with a variety of methods, the following strategy produces consistent good quality (low residual) stacked data {(note, we assume both the short-wavelength filter image we wish to fix and the long-wavelength filter reference image are at least pedestal background subtracted)}:

{
\begin{enumerate}[start=0]
\item identify a target short-wavelength filter image that has wisps that needs to be removed ($S_{\text{in}}$); B) identify a long-wavelength filter (cosmetically F444W is often best, but F356W works similarly well) stacked mosaic covering a similar region of sky ($L_{\text{in}}$);
\item project $L_{\text{in}}$ to match the world coordinate system (WCS) of $S_{\text{in}}$, this projected image will be $L_{\text{warp}}$;
\item calculate the most extreme plausible blue colour ($B_{\text{lim}}$) that the user might expect between the short and long filters (this could be computed theoretically from e.g.\ the spectrum of OB stars, or estimated dynamically from the images being processed, see below);
\item create a map of the flux differences between the long and short filter frames -- this is the initial wisp template $WT$, where $WT = S_{\text{in}} - (L_{\text{warp}} \times B_{\text{lim}})$;
\item mask pixels (i.e.\ set to NA or NaN) in $WT$ that are above some `clip' threshold (this avoids bright pixels around stars being flagged as wisp features purely due to wavelength-dependent differences in point spread functions): $WT[WT > \text{clip}] = \text{NA}$;
\item smooth $WT$ with a kernel that is smaller than the highest frequency wisp, but large enough to overcome Poisson photon noise in the map;
\item set negative valued pixels in $WT$ to a masked value (e.g. NA or NaN): $WT[WT < 0] = \text{NA}$;
\item smooth $WT$ with a low frequency kernel to make $WT_{\text{lo}}$;
\item replace masked values and lower valued pixels in $WT$ with the corresponding $WT_{\text{lo}}$ pixels: $WT[\text{NA} \mid WT < WT_{\text{lo}}] = WT_{\text{lo}}[\text{NA} \mid WT < WT_{\text{lo}}]$;
\item subtract the final $WT$ map from the target frame to be corrected: $S_{\text{out}} = S_{\text{in}} - WT$.
\end{enumerate}
}

Regarding step 2 above, we found that dynamically calculating the bluest plausible colours ($B_{\text{lim}}$) based on image statistics works well in practice, and lends the technique greater flexibility when using different combinations of short- and long-wavelength filters. There are a number of ways to achieve this reasonably, but our preferred strategy is:

{
\begin{enumerate}[start=2]
\item
    \begin{enumerate}
    \item create a map of WCS-matched relative fluxes ($RF$) between $S_{\text{in}}$ (numerator) and $L_{\text{warp}}$ (denominator);
    \item select pixels in $RF$ corresponding to the brightest 95\% of pixels in the long-wavelength stacked WCS-matched reference image (these pixels are highly likely to belong to real sources given the depth of PEARLS and similar imaging data);
    \item remove any negative-valued pixels (there should not be many, but we only want to analyse positive-ratio pixels);
    \item compute the $95^{th}$ percentile of the remaining $RF$ pixels (ranked low to high), and treat this as the bluest plausible value ($B_{\text{lim}}$) for future analysis (this is the main parameter that controls how aggressively sources are masked out, higher meaning more masking).
    \end{enumerate}
\end{enumerate}
}

\begin{figure*}[t]
\begin{center}
\includegraphics[width=5in]{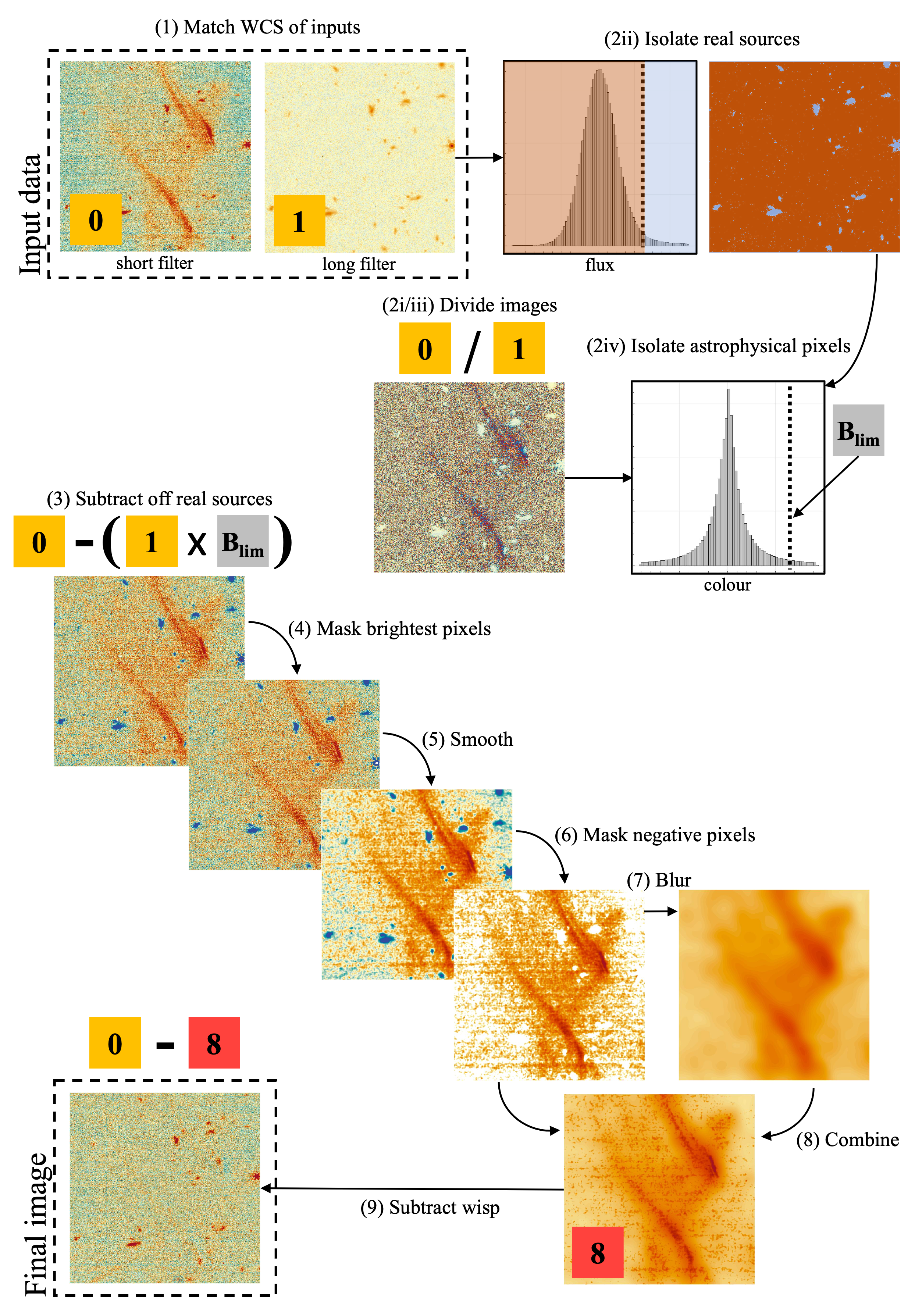}
\caption{Schematic view of the processing steps required to remove wisps. These numbered steps correspond those outlined in Section \ref{sec:process}.}
\label{fig:workflow}
\end{center}
\end{figure*}


{
Figure \ref{fig:workflow} offers a visual schematic overview of the main processing steps outlined above, making it clearer what the typical inputs and outputs of each step should be. The main parameters discussed above were chosen based on visual inspection of wisp-subtracted frames and the results of mock wisp-removal simulations (discussed in more detail later in Section \ref{calibrate}). The quality of wisp removal is not highly sensitive to the exact values chosen. For the source threshold, values between 80--99\% of brightest pixels work well (we default to 95\%). For defining blue colours, values between 80--99\% work well (we default to 95\%). For final clipping, values between 99--100\% work well (we default to 99.7\%). The optimal smoothing is more subjective, but for PEARLS-depth data, setting the standard deviation of a 2D Gaussian smoothing kernel to 2 pixels appears to be a good compromise of resolution and signal, but values between 0.5 and 3 pixels work similarly well.
}


{
We also investigated the best strategy for defining wisp templates in the regions where we have masked objects. In general, the complex structure of the wisps cannot be perfectly approximated by interpolating through large regions of missing pixels. However, combining both the higher frequency smoother data with a lower frequency representation that provides values in masked regions proved to be the best option (based on mock wisp removal simulations, as we discuss later in this work). In particular, it is better to construct a combined high and low frequency wisp template than merely masking the source positions entirely. Note, the primary aims of removing wisps is to produce better (flatter) backgrounds for later removal, and also to remove completely spurious sources produced by wisps. For sensibly dithered data (at the scale of the wisps and larger than most sources of interest) clipping and/or median stacking should mitigate any small residual wisp signal lying within sources.
}

\subsection{Calibrating and Testing}
\label{calibrate}

It is important to test and calibrate the aggressiveness of any wisp removal scheme, since we do not want to remove real flux from the image or (ideally) leave any wisp flux behind. To optimise our wisp removal parameters we used a median-stacked wisp template for filter F200W and detector NRCB4 since this is generally the worst wisp-affected filter-detector combination (see Figure \ref{fig:wisp_process_mock}). We then embedded this wisp template in 64 NRCA1, NRCA2, NRCB1 and NRCB2 frames (which are visually wisp-free) and attempted to remove the introduced wisp with our automated scheme as outlined above.

Depending on the frame (and the fluxes of the real sources) the added wisp contains in the region of 10-20\% of the total pedestal subtracted flux in the image. It is often the brightest single `object' in the image, but since it is distributed over a large area it is usually also the lowest average surface brightness structure. These properties give some sense of the significance of wisps in terms of how much they can compromise JWST data, and also why removing them is a challenging task.

\begin{figure}[]
\begin{center}
\includegraphics[width=5in]{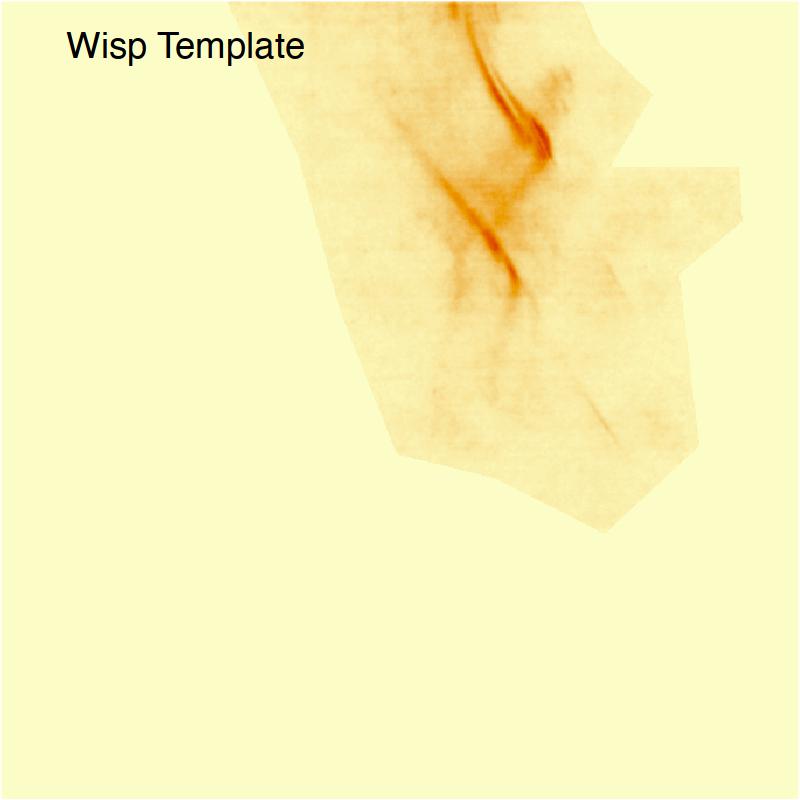}
\caption{Example of a simulated wisp template for a F200W filter frame and detector NRCB4. This is a masked, pixel-clipped and slightly smoothed (to reduce pixel noise) version of the bottom right panel of Figure \ref{fig:STScI_wisp_templates} using the wisp templates made available at STScI (linked above). Colour scaling is per Figure \ref{fig:STScI_wisp_templates}.}
\label{fig:wisp_template}
\end{center}
\end{figure}

The default values presented above allow for nearly optimal wisp removal, where we err on the slightly aggressive side since local regions of over-subtraction are more easily fixed during later image stacking (the worst over-subtractions happen near bright stars where pixels will be flagged). Figure \ref{fig:wisp_process_mock} shows an example of a frame where the embedded wisp has been cleanly removed, and the final wisp template generated is visually close to the unknown (to the algorithm) wisp template embedded. The highest-resolution features of the embedded wisp cannot be fully recovered due to the lower signal-to-noise of the real data in a single frame.

\begin{figure}[]
\begin{center}
\includegraphics[width=2in]{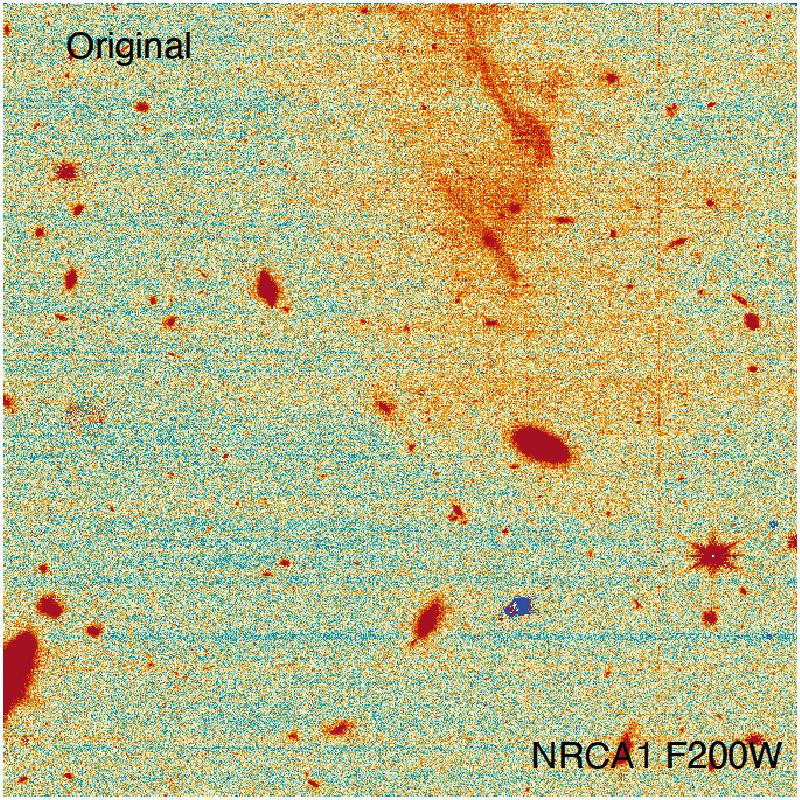}
\includegraphics[width=2in]{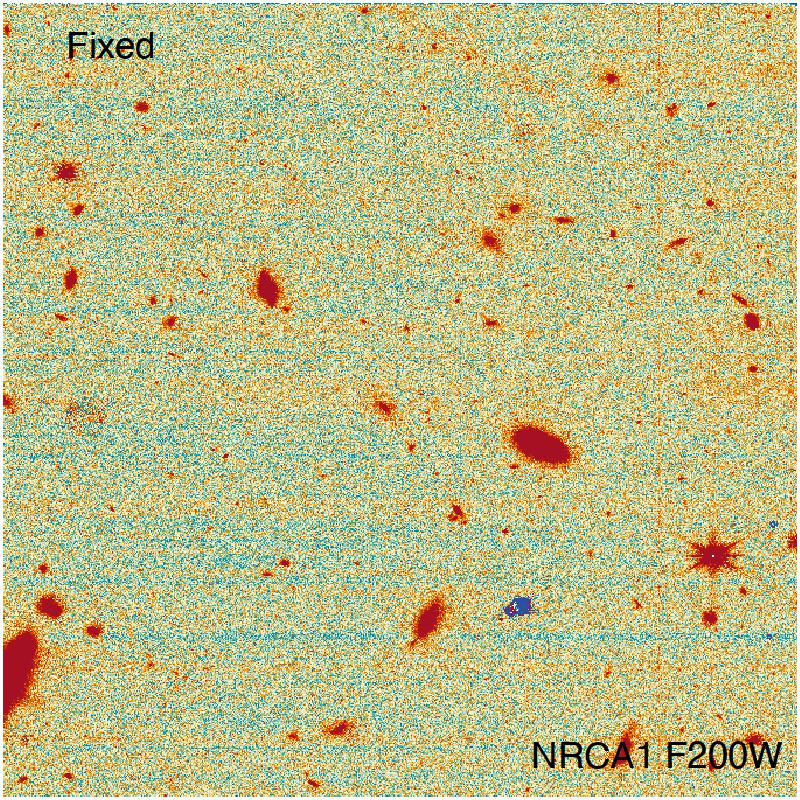}
\includegraphics[width=2in]{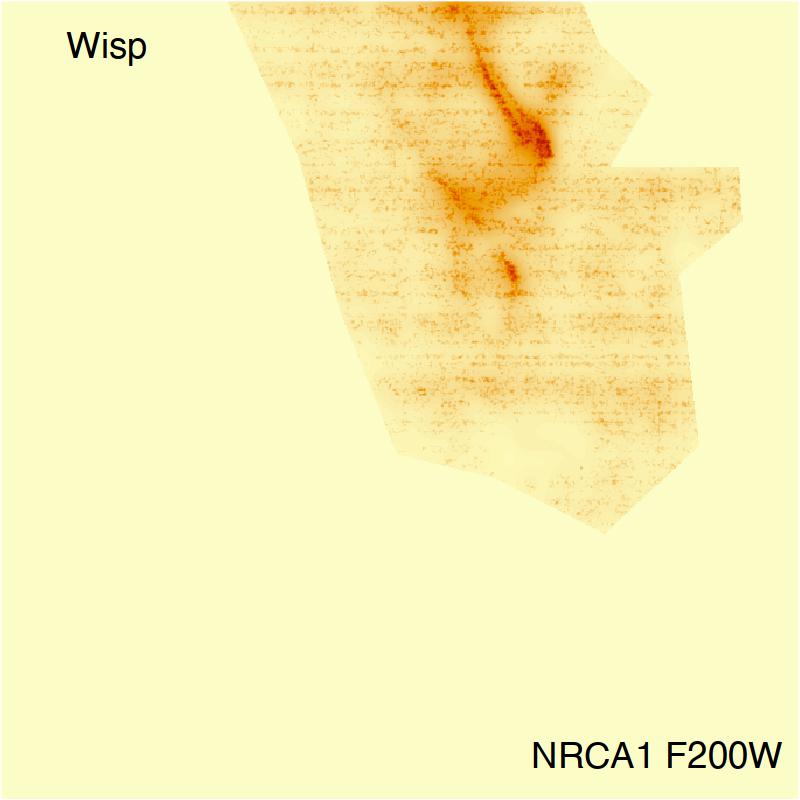}
\caption{Example of simulated wisp removal. The wisp template shown in Figure \ref{fig:wisp_template} is embedded into the image at the level seen for particularly bad wisps. The image on the left is a F200W filter frame from NRCA1 taken as part of PID 2738 (VID 2738008001). The original image was visually free of wisps before the template was added. The centre image shows the result of the wisp subtraction method, where the cloudy structure at the top middle of the left panel has largely been removed. The right hand image shows the wisp template that was subtracted (white pixels are the fully masked regions of the wisp that correspond to pixels we do not want to modify). Colour scaling is per Figure \ref{fig:STScI_wisp_templates}.}
\label{fig:wisp_process_mock}
\end{center}
\end{figure}

\begin{figure}[]
\begin{center}
\includegraphics[width=5in]{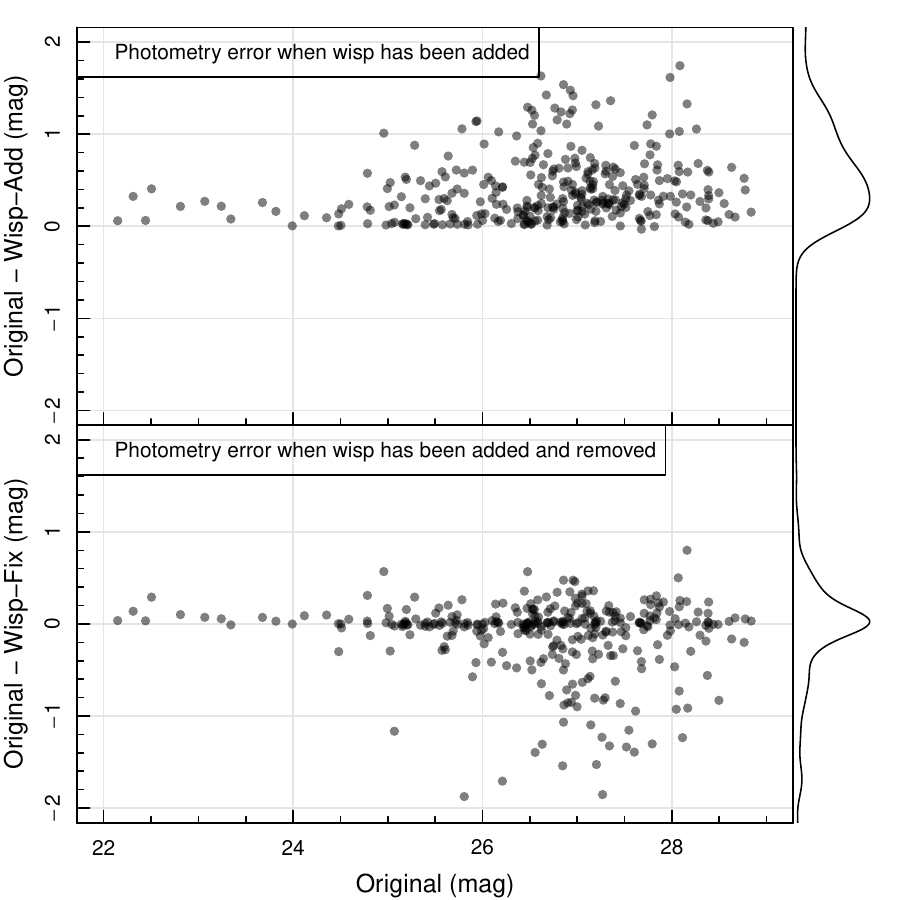}
\caption{The impact of wisps on photometry. Here we ran \profound{} on the example mock data presented in Figure \ref{fig:wisp_process_mock} and limit the results to within the wisp polygon region. The top panel shows the magnitude difference between the original (wisp-free) frame and the mock frame (wisp-added). The bottom panel shows the magnitude difference between the original (wisp-free) frame and the processed frame (wisp-fixed). The wisp-added data clearly has highly compromised photometry, with sources appearing appreciable brighter due to the presence of the wisp. The wisp-fixed frame still has some scatter, but the photometry is now median unbiased (0.298 mag bright with the wisp, and 0.003 mag faint after processing), and significantly improved in most respects.}
\label{fig:wisp_mag_improve}
\end{center}
\end{figure}

{
Taking the example shown in Figure \ref{fig:wisp_process_mock}, we ran \profound{} on the original (wisp-free) frame, the mock frame (wisp-added) and the processed frame (wisp-fixed). This allows us to asses the impact that wisps have on our basic photometry, and confirm that the wisp processing methodology outlined above does more good than harm. Figure \ref{fig:wisp_mag_improve} presents the main results, where we clearly see the photometry in the wisp-added frame is highly compromised (sources are systematically too bright throughout). In comparison, the wisp-fixed frame is close to having unbiased photometry, and would be much preferred. We note that this particularly simulation would be quite an extreme combination of faint sources sitting on top of a very pronounced wisp, so typical PEARLS data would not need such extreme processing to remove the wisp. Further, a sensible dither and stacking strategy would make full use of the median unbiased qualities of the wisp removal presented here; i.e.\ dithers should be larger than typical source sizes and large enough to avoid the brightest parts of the wisp.
}

\begin{figure}[]
\begin{center}
\includegraphics[width=5in]{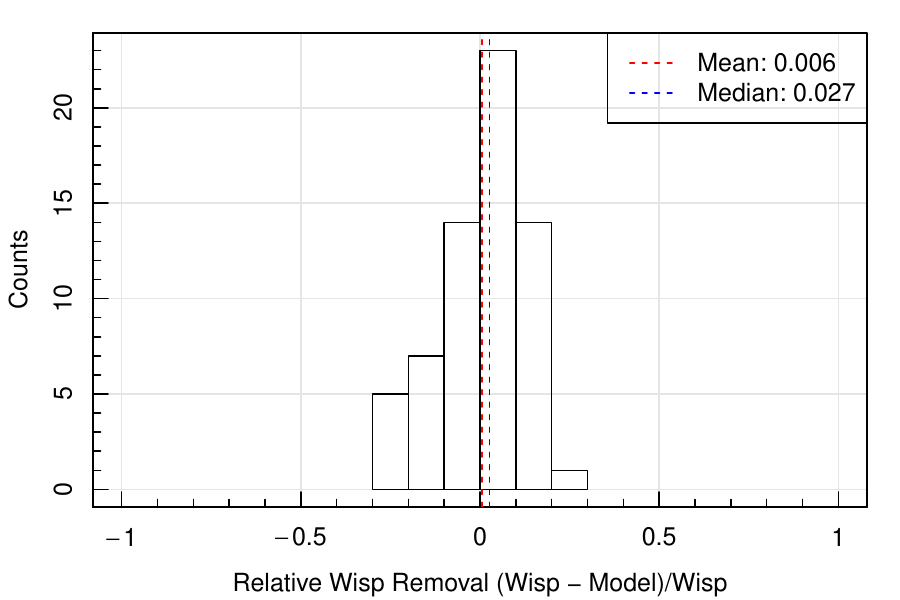}
\caption{Relative wisp residual statistics. Negative values indicate frames where our wisp removal method has over-subtracted compared to the original embedded wisp template, and positive values indicate where we have under-subtracted. The mean and median statistic suggests a small degree of systematic over-subtraction on average (close to 0). The maximum over-subtraction is less than 20\% of the wisp signal, and the maximum under-subtraction is less than 40\% of the wisp signal (i.e. we err on the side of not being too aggressive with our removal process).}
\label{fig:wisp_stats}
\end{center}
\end{figure}

{Now we consider the main outcomes for all 64 simulated and processed wisp frames.} Overall the parameters discussed above work well for JWST NIRCam data, with on average {99.4\%} (mean) of the wisp template flux accounted for during this automatic modelling process. Figure \ref{fig:wisp_stats} presents the overall statistics for 64 simulated wisp frames. We find that whilst we do not, on average, perfectly remove the embedded wisps, our processing technique always improves the flux measurements (all values are bounded between -1 and +1), i.e.\ we do more good than harm in all situations. Since the higher surface brightness parts of the wisp are usually very well removed, the remaining features (be they slightly positive or negative) are much smoother and easier to remove at later stages of the PEARLS processing pipeline. In particular the $1/f$ and \profound{} sky subtraction stages work together to flatten out the remaining residuals in most situations.

\subsection{Achieving $\sqrt{T}$ Stacking}

As a final test, a comparison was made of images stacked using \propane\footnote{https://github.com/asgr/ProPane}. The wisp-free stack of F200W data was compared to the wisp-added stack (using the mock template above), and then the wisp-added and corrected stack (`wisp-fixed', using the approach discussed in this work). {In this test all frames had the same exposure time (419s), so a $\sqrt{T}$ depth increase is equivalent to a $\sqrt{N_{frame}}$ relative decrease in the sky noise.} For the NEP-TDF data used for the template simulations above, the deepest regions of our stack combine as many as eight individual frames (10.7\% of the total area), and due to the masking and dithering strategy used there are many pixels with only one contributing frame (1.6\%). At nominal depth (most of the area), four frames contribute to the stack.

To test the impact of stacking, we first create a deep segmentation map using the cosmetically clean F444W data warped (reprojected and resampled) with \propane{} to the same WCS as our target stack. To do this we run \profound{} with default settings \cite[for details see][]{2018MNRAS.476.3137R}. The non-object pixels identified during this process can then be used to measure background statistics, and to confirm whether we are able to achieve $\sqrt{T}$ noise reduction with our wisp removal strategy.

Firstly, we find that using the wisp-free data we achieve close to a $\sqrt{T}$ increase in image depth (reduction in background root-mean-square noise; RMS). More precisely, we find our image depth increases as $\frac{\sqrt{T}}{1.012^T}$. The slight deviance from perfect $\sqrt{T}$ noise reduction is mostly due to covariance created during stacking (which will always be present when combining non pixel-aligned data) and the presence of non-detected sources in the `background'. In addition, absolute read-noise levels will prevent $\sqrt{T}$ noise reduction. This analysis is used as our best-case-scenario reference measurement for wisp removal.

\begin{figure}[]
\begin{center}
\includegraphics[width=2in]{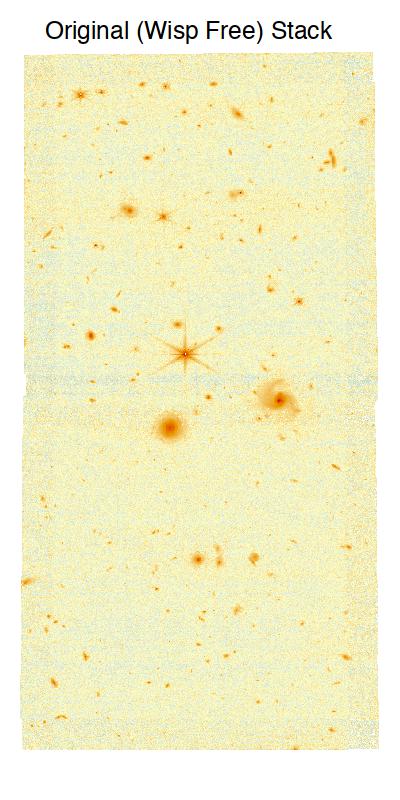}
\includegraphics[width=2in]{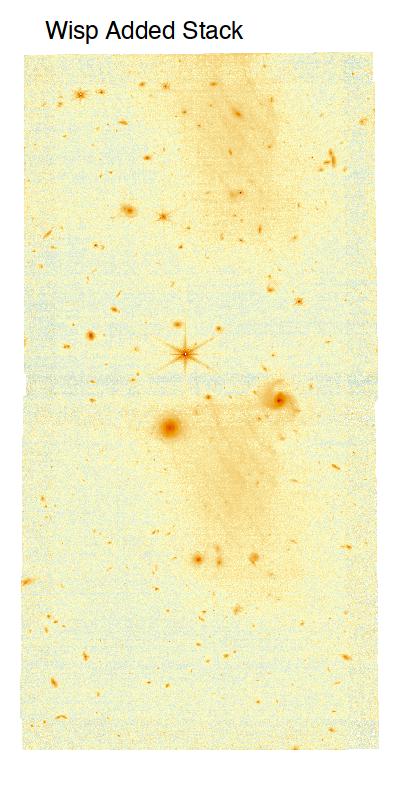}
\includegraphics[width=2in]{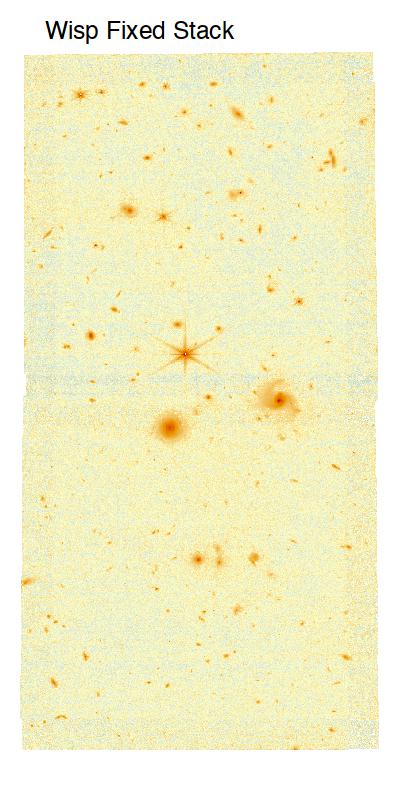}
\caption{Example of simulated wisp stacking. The left panel shows a stack of 16 frames (with maximally eight frames overlapping) that have no wisps present (Original). The middle-panel shows the same stack but using the mock data with a wisp template added (wisp added). The right panel shows the same stack, but applying our wisp removal algorithm to the wisp-added data (wisp fixed). Colour scaling is per Figure \ref{fig:STScI_wisp_templates}.}
\label{fig:stack_mock}
\end{center}
\end{figure}

We then compare the background RMS using the wisp-added data, and then finally the wisp-fixed data. The stacked images for all three scenarios are shown in Figure \ref{fig:stack_mock}. When measuring the background statistics of the wisp-added data we find an RMS that is a factor {1.474} larger over the whole stacked frame (for pixels with eight frames contributing, our maximum depth). This is akin to saying we reduce the exposure time by a factor 2.2. When applying our wisp-removal strategy we obtain a background RMS that is a factor {1.015} larger over the same pixels, which is an effective reduction in exposure time of only a factor {1.031}. Using the above slight non-linearity with $\sqrt{T}$, we find that we measure an RMS of our deepest pixels (eight frames contributing) a factor 1.10 larger than the $\sqrt{8}$ reduction we could expect with no data limitations. This means the limiting effects when increasing the image depth (with dithered exposures) is dominated by covariance and un-detected sources, and not the wisps (when using our wisp-removal process).

\section{Results}

The initial implementation of this processing method is written in \R{} \cite{R-Core-Team:2021wf} since many of the routines necessary are provided by the \protools{} suite of packages written by ASGR\footnote{https://github.com/asgr/ProTools}, and several other processing steps in the PEARLS pipeline also use this software. The code for the main processing steps is provided in the Appendix, where we assume that it should be conceptually easy to re-write the procedure into any desired target language starting from this \R{} code.

\begin{figure}[]
\begin{center}
\includegraphics[width=2in]{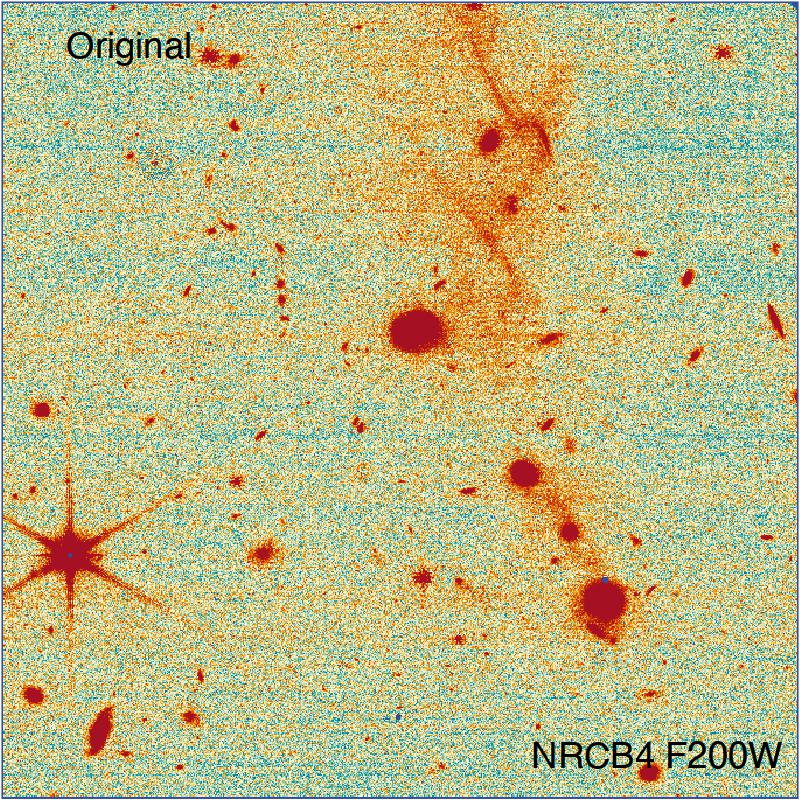}
\includegraphics[width=2in]{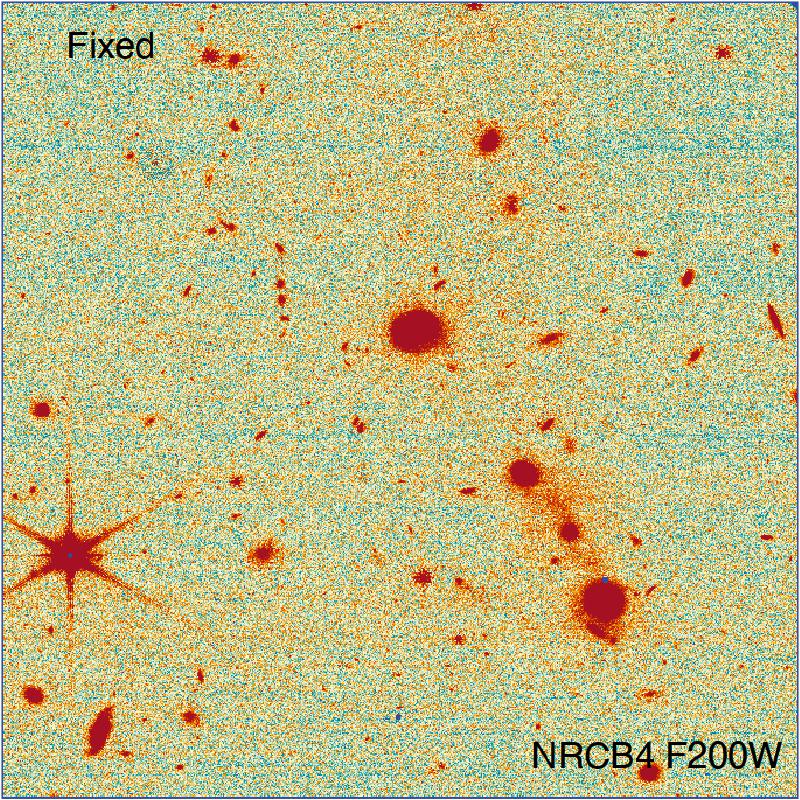}
\includegraphics[width=2in]{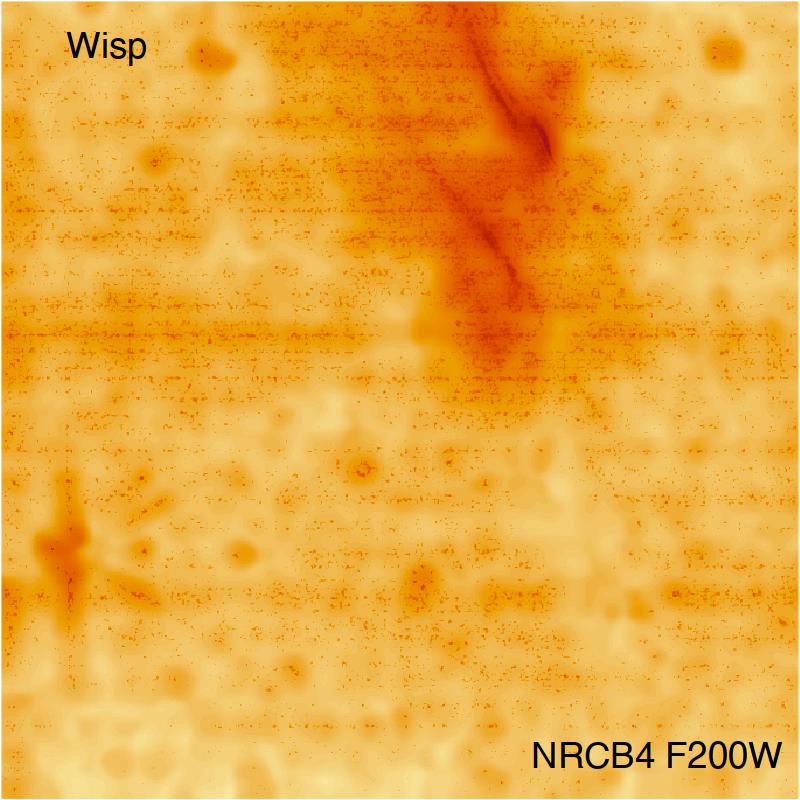}
\caption{Example of real wisp removal. The image on the left is a F200W filter frame from NRCB4 taken as part of PID 2738 (VID 2738002001). The centre image shows the result of the wisp-subtracted image where the cloudy structure in the top middle of the frame has largely been removed. The right-hand image shows the derived wisp template that was subtracted. Colour scaling is per Figure \ref{fig:STScI_wisp_templates}.}
\label{fig:wisp_process}
\end{center}
\end{figure}

We applied the wisp-removal process to the worst-affected public PEARLS data in the NEP-TDF (PID 2738, VID 2738002001). This was done at {\it cal} stage 2, i.e.\ the output of {\sc calwebbimage2}, which is the step immediately before any more aggressive $1/f$ read-out noise removal in our pipeline. Figure \ref{fig:wisp_process} shows the main flow of the pipeline processing, with the original wisp-compromised image in F200W shown on the left, and the removed (middle) and final wisp template (right). Visually and statistically (in terms of the sky characteristics) we find this processing method has hugely improved the quality of the data (achieving a level similar to the tests described above).

\subsection{Effect of Varying the Filter of the Same Detector}

\begin{figure}[]
\begin{center}
\includegraphics[width=2in]{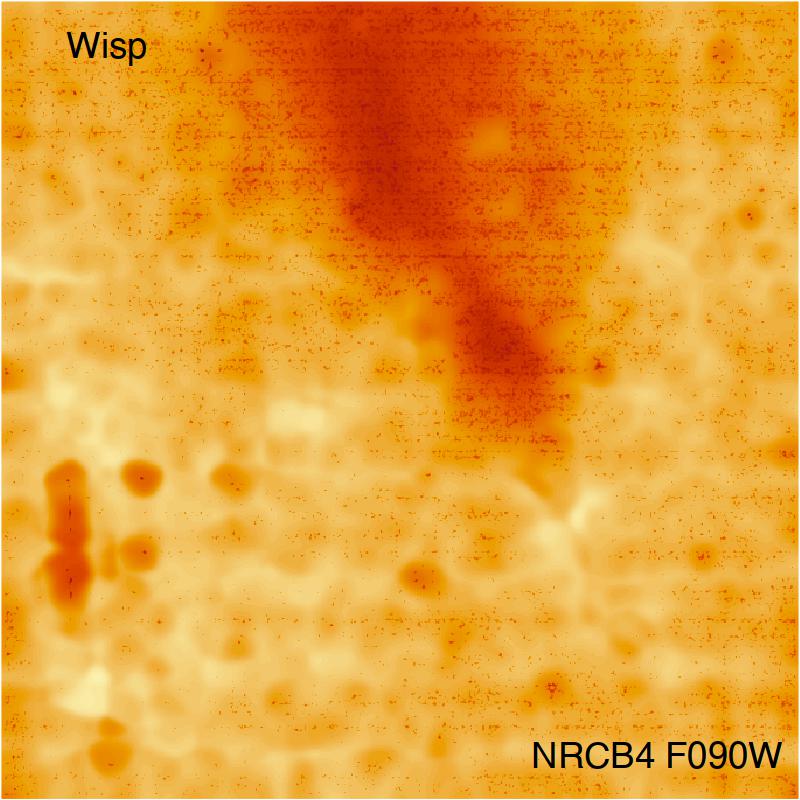}
\includegraphics[width=2in]{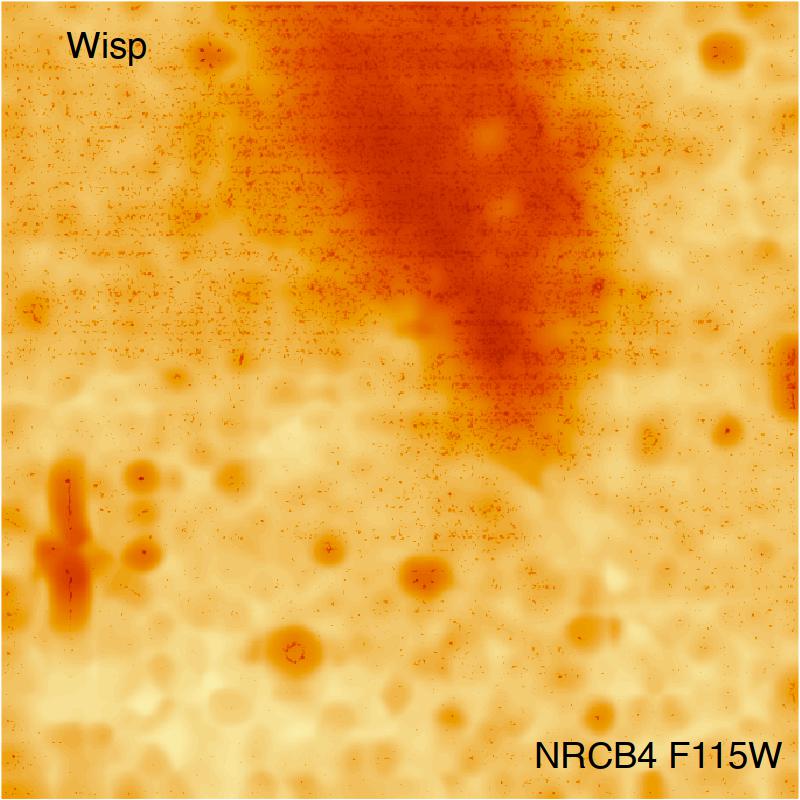}
\includegraphics[width=2in]{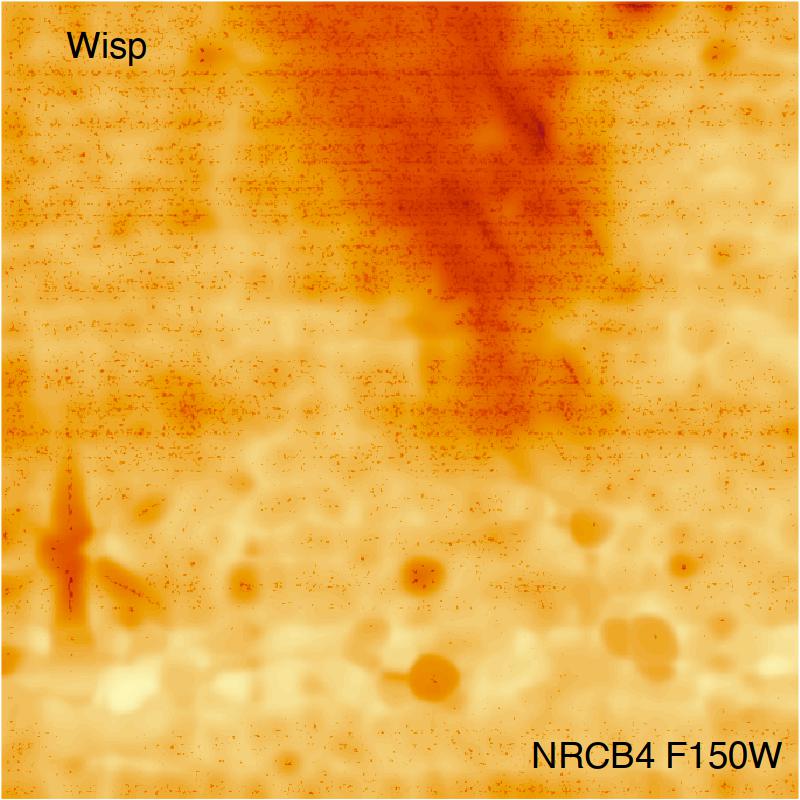}
\caption{The same Visit ID as presented in Figure \ref{fig:wisp_process} but with different filters. Left image is F090W, middle image is F115W, right image is F150W. Colour scaling is per Figure \ref{fig:STScI_wisp_templates}.}
\label{fig:vary_filter_static_detector}
\end{center}
\end{figure}

Universally across filters, detector NRCB4 is the worst affected by wisps. The wisps tend to appear in broadly the same region, but the resolution and surface brightness of the wisp varies with the  filter wavelength. This can be seen in Figure \ref{fig:vary_filter_static_detector} where we present the derived wisp template for the same region of sky but in different filters. F150W shows some evidence of the higher surface brightness cirrus structure in the upper right of the wisp seen clearly in Figure \ref{fig:wisp_process} (right panel). F090W and F115W have much more extended wisp structures, and both lack the sharper higher surface brightness diagonal features seen in F200W.

\subsection{Effect of Small Dithers for the Same Filter and Detector}

\begin{figure}[]
\begin{center}
\includegraphics[width=2in]{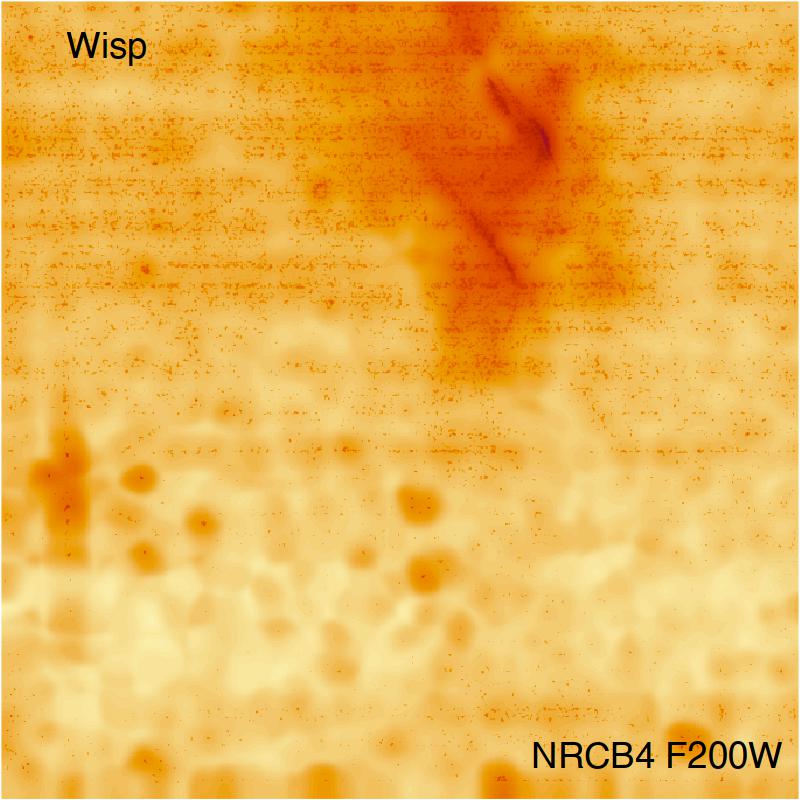}
\includegraphics[width=2in]{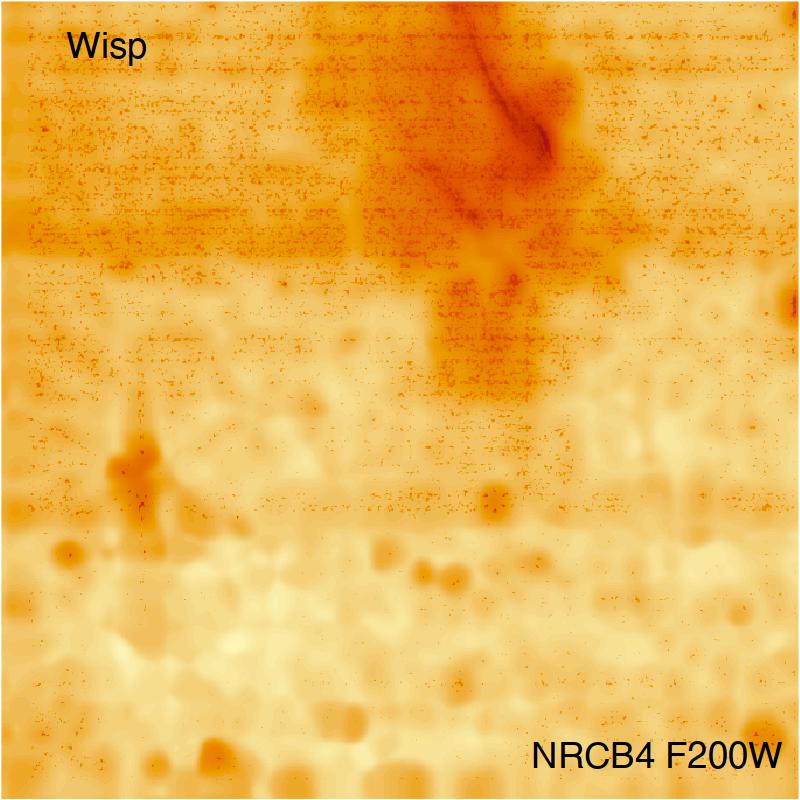}
\includegraphics[width=2in]{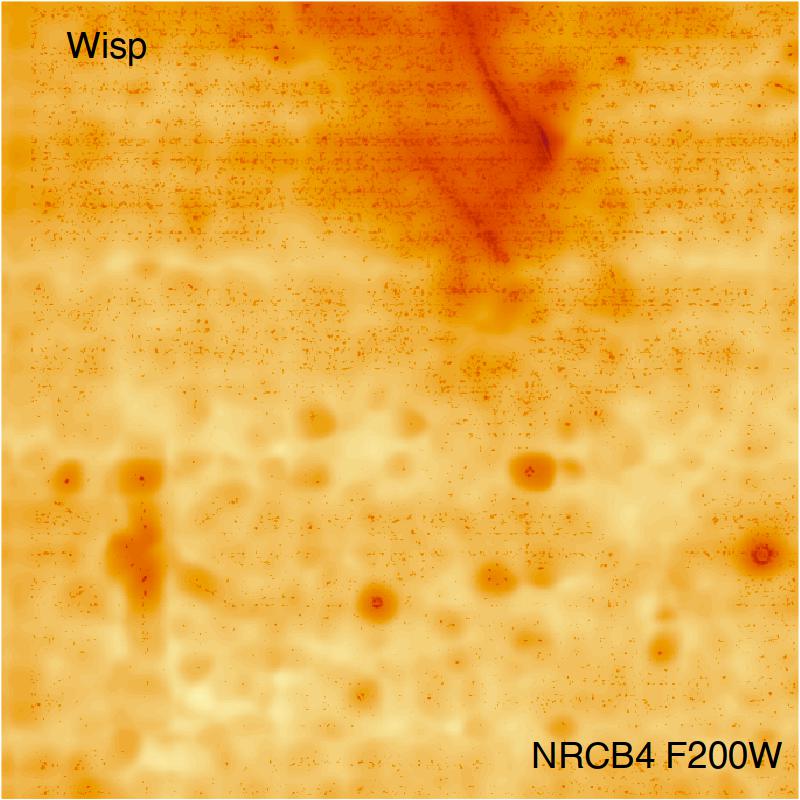}
\caption{The same Visit ID as presented in Figure \ref{fig:wisp_process} and same F200W filter, but showing the three subsequent dithers within the same NRCB4 detector. Colour scaling is per Figure \ref{fig:STScI_wisp_templates}.}
\label{fig:static_filter_static_detector}
\end{center}
\end{figure}

A key question is how much our derived wisp template varies with small dithers, i.e.\ when we are considering a fixed filter and the same detector. Since F200W in detector NRCB4 is the most dramatically affected by wisps, we consider small dithers of the data presented in Figure \ref{fig:wisp_process}. Figure \ref{fig:static_filter_static_detector} shows the effect of such small (a few arc second) dithers. The general shape is quite consistent, but in detail it is clear the intensity of the higher and lower surface brightness parts of the wisp do vary quite notably.

\begin{figure}[]
\begin{center}
\includegraphics[width=2in]{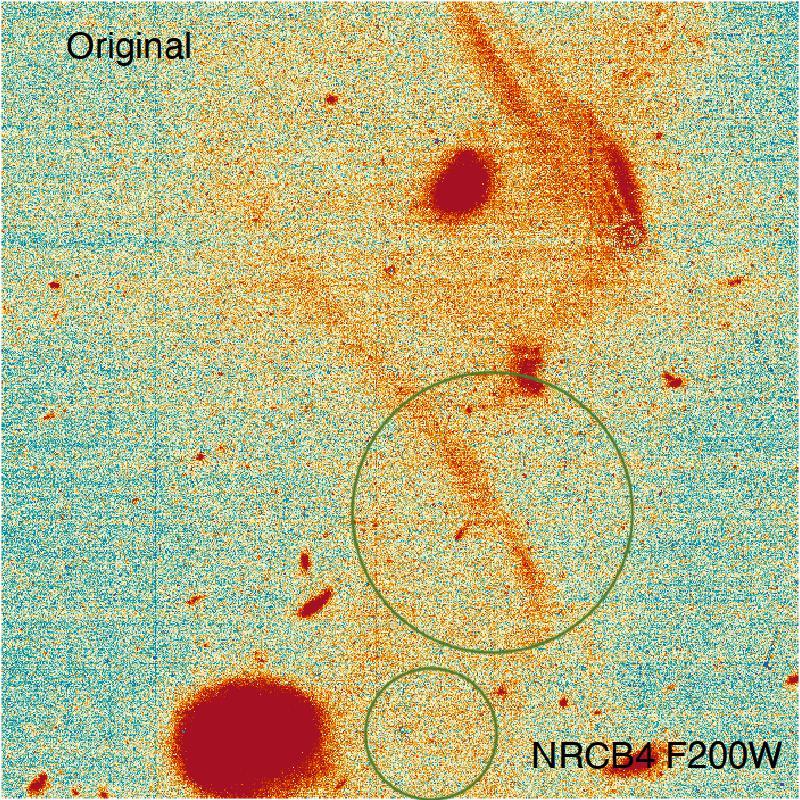}
\includegraphics[width=2in]{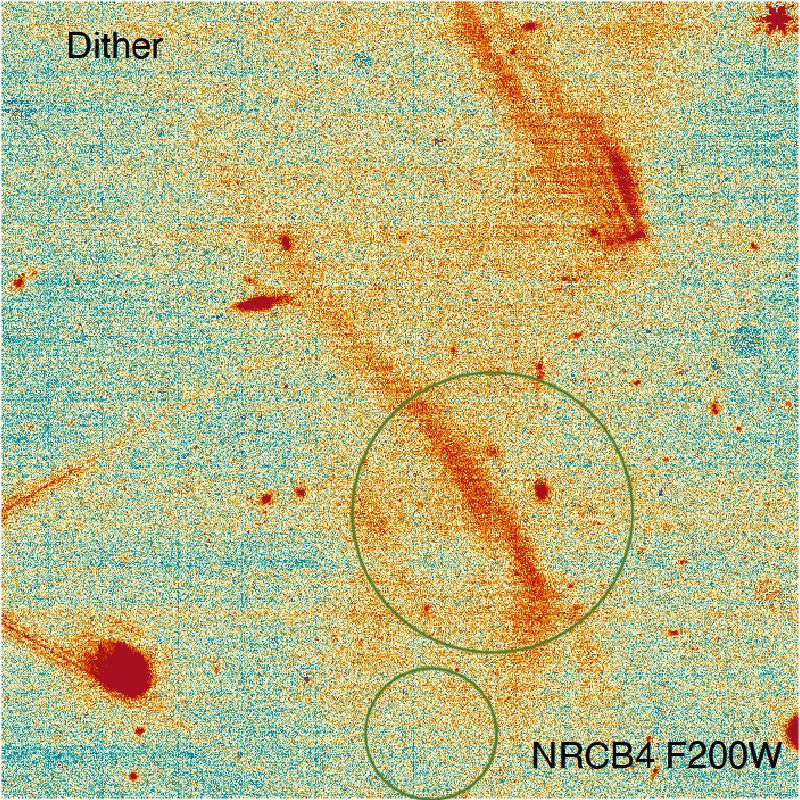}
\includegraphics[width=2in]{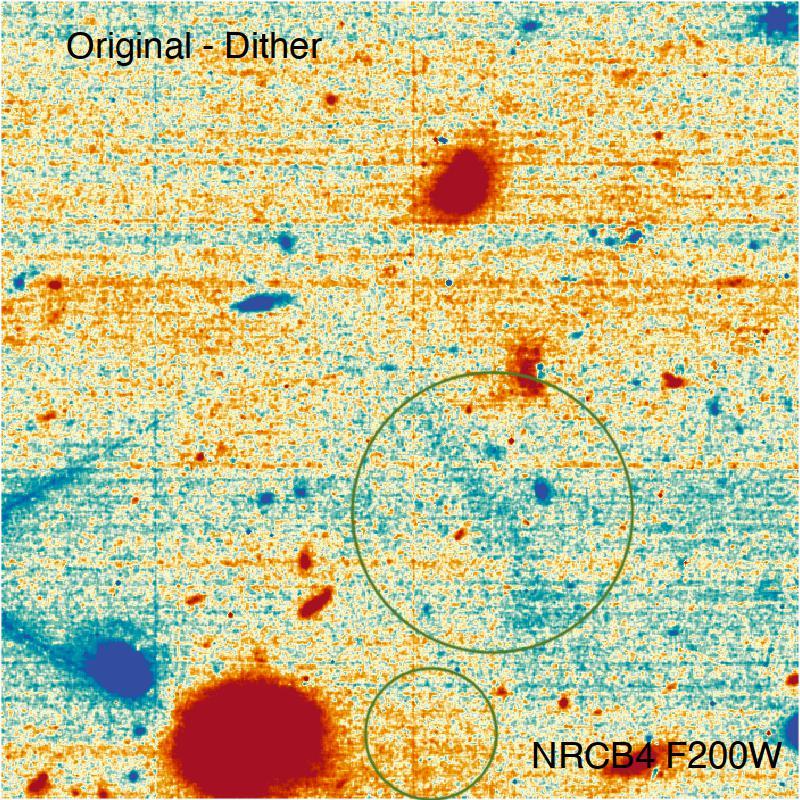}
\caption{A zoomed in view of the effect of wisps varying with dithers for the F200W filter and NRCB4 detector. Colour scaling is per Figure \ref{fig:STScI_wisp_templates}. Left and Middle image are two dithers, the right image is the difference. The green circles highlight regions where the wisps appear to vary the most between dithers.}
\label{fig:static_filter_static_detector_zoom}
\end{center}
\end{figure}

{
To highlight the effect of wisps varying between dithers even for the same detector/filter combination within the same program visit, Figure \ref{fig:static_filter_static_detector_zoom} shows the wisps and the image differences for a zoomed in region of the most prominent wisp we find in our PEARLS data (top middle of F200W filter and NRCB4 detector). The biggest differences are highlighted with green circles, where a mixture of lower surface brightness features and a prominent diagonal wisp are seen to differ between dithers.
}

\subsection{Effect of the Same Filter and Varying Detectors}

\begin{figure}[]
\begin{center}
\includegraphics[width=2in]{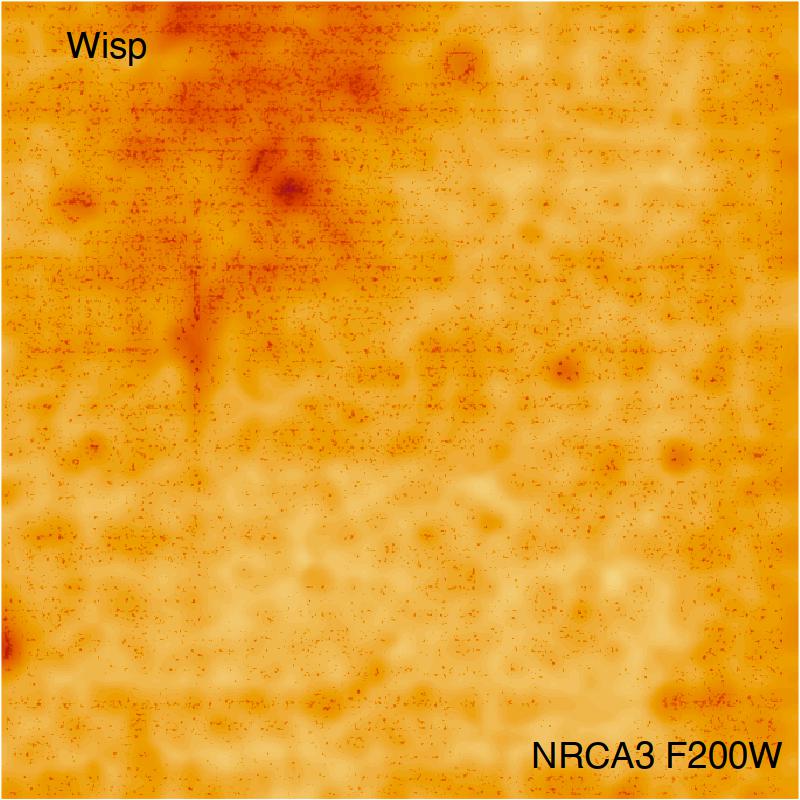}
\includegraphics[width=2in]{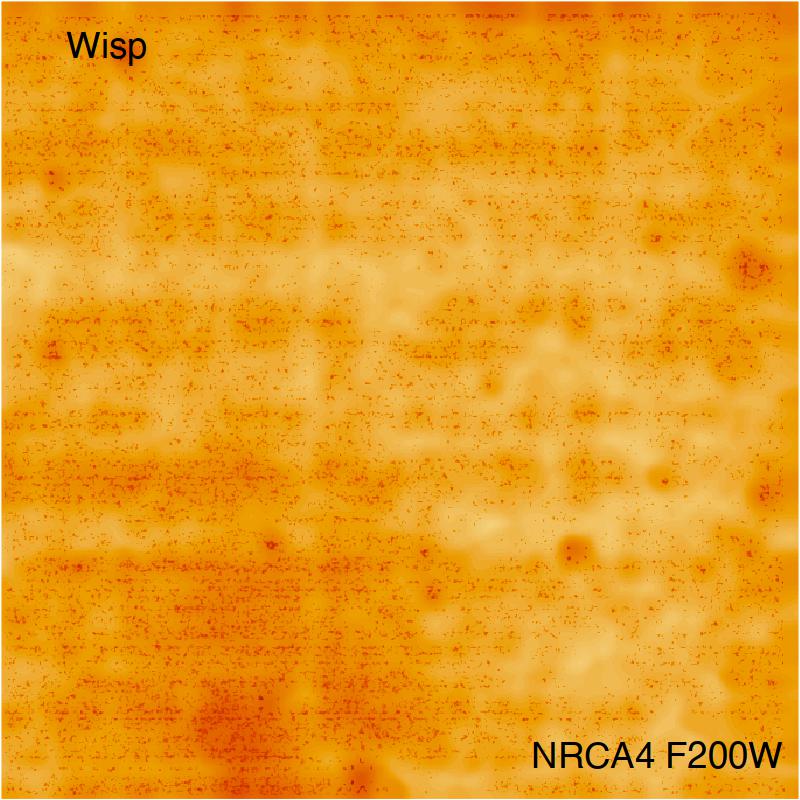}
\includegraphics[width=2in]{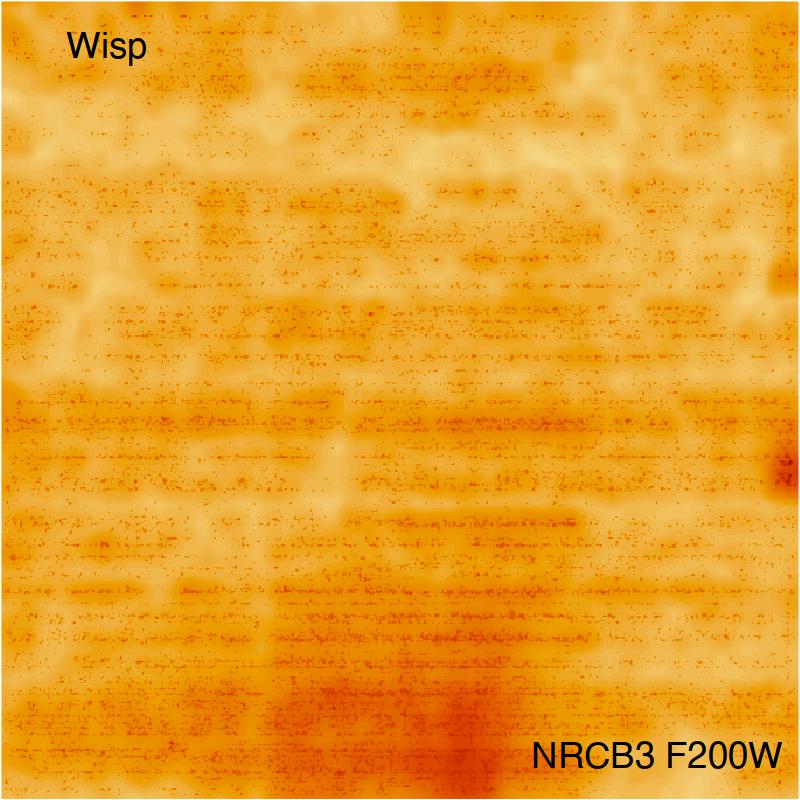}
\caption{The same Visit ID as presented in Figure \ref{fig:wisp_process} and same F200W filter, but showing the NRCA3 (left), NRCA4 (middle) and NRCB3 (right) detectors, each of which is less affected by wisps than NRCB4. Colour scaling is per Figure \ref{fig:STScI_wisp_templates}.}
\label{fig:static_filter_vary_detector}
\end{center}
\end{figure}

As discussed above, the NRCB4 is the detector most seriously affected by wisps in general. This is followed by the NRCB3 detector (where they tend to present in the bottom part of the image) and to a much lesser extent NRCA3 and NRCA4. This can be seen in Figure \ref{fig:static_filter_vary_detector}, where there is some mild wisp structure towards the bottom of NRCB3 (right panel) and much less evident structure in NRCA3 and NRCB4 (left and middle panels respectively). In this example, removing wisps from NRCA3 and NRCA4 is probably not necessary, but the method used ensures that we will, on average, correct the effect of wisps more than remove any real flux (i.e\ Figure \ref{fig:wisp_stats}).

\section{Conclusions}

In this paper we have presented a pragmatic, dynamic strategy to mitigate wisps in NIRCam short-wavelength detector data. This only relies on a good reference stack in one of the long-wavelength NIRCam detectors, and \emph{does not require the long-term construction of wisp templates}. This is significant, since it is clear after 9 months of science operations that the structure and strength of wisps is highly sensitive to the exact position and orientation of JWST. These wisps presumably have a scattered light origin, where the presence and spectrum of off-axis bright stars has a complex effect on the wisp pattern found in the raw images. Our hope is that many medium and deep surveys with JWST will benefit from incorporating a similar wisp-removal strategy, allowing survey depth to improve with $\sqrt{T}$ throughout the image.

We include reference code written in the \R{} programming language in the Appendix. All the functions and algorithmic steps can be easily replaced with {\sc Python}, {\sc IDL} or {\sc Julia} equivalents.

\section*{Acknowledgements}

ASGR acknowledges funding by the Australian Research Council (ARC) Future Fellowship scheme (FT200100374, `Hot Fuzz'). RAW and RAJ acknowledge support from NASA grants NAG5-12460, NNX14AN10G and 80NSSC18K0200 from GSFC. SPD acknowledges funding by the Australian Research Council (ARC) Laureate Fellowship scheme (FL220100191). CNAW acknowledges support from the NIRCam Development Contract NAS5-02105 from NASA Goddard Space Flight Center to the University of Arizona. {We thank Prof. John Peacock for a useful suggestion to consider the bluest low-SB features that may be a real astrophysical source. We thank the anonymous referee for providing constructive comments that improved aspects of the core algorithm and the comprehensibility of the paper.}
\newline

\noindent
\emph{Software:} \profound: https://github.com/asgr/ProFound \cite{2018MNRAS.476.3137R, 2018ascl.soft04006R} (LGPL-3). \propane: https://github.com/asgr/ProPane (LGPL-3). \Rfits: https://github.com/asgr/Rfits (LGPL-3). \Rwcs: https://github.com/asgr/Rwcs (LGPL-3).
\bibliographystyle{iopart-num}
\bibliography{wisp}

\appendix

\section{R Code Implementation}

Below is a compact example of our wisp removal scheme written in \R. Note you will need the \profound, \propane, \Rfits{} and \Rwcs{} packages also (all available on GitHub at user \emph{asgr}). These steps correspond to those presented in Section \ref{sec:process} and the schematic overview in Figure \ref{fig:workflow}.

\begin{alltt}
library(ProFound)
library(ProPane)
library(Rfits)
library(Rwcs)

wisp_im = Rfits_read_image(`wisp_im.fits') \textcolor{cyan}{#Read FITS file}
ref_im = Rfits_read_image(`ref_im.fits') \textcolor{cyan}{#Read FITS file}

wisp_fix = wispFixer(wisp_im, ref_im) \textcolor{cyan}{#Run wisp removal function}

Rfits_write_image(wisp_fix, `wisp_fix.fits') \textcolor{cyan}{#Write FITS file}

\textcolor{cyan}{#Where wispFixer is the following function:}
wispFixer = function(wisp_im, \textcolor{cyan}{#Rfits short filter image with wisps}
                     ref_im, \textcolor{cyan}{#Rfits long filter reference image}
                     source_threshold = 0.95, \textcolor{cyan}{#Threshold on reference image}
                     scale_threshold = 0.95, \textcolor{cyan}{#Quantile to define blue things}
                     clip_threshold = 0.997, \textcolor{cyan}{#Avoid very bright pixels}
                     sigma_hi = 2 \textcolor{cyan}{#High freq. wisp smoothing kernel},
                     sigma_lo = 20 \textcolor{cyan}{#Low freq. wisp smoothing kernel}
)\{
\textcolor{cyan}{#Note: wisp_im and ref_im are assumed to be background subtracted}

\textcolor{cyan}{#Step 1: Warp ref_im to match wisp_im WCS}
    ref_im_warp = propaneWarp(ref_im, header_out = wisp_im$raw)
 
\textcolor{cyan}{#Step 2i: Create ratio image}
    relflux = (wisp_im$imDat / ref_im_warp$imDat)
    
\textcolor{cyan}{#Step 2ii: Select the pixel threshold of real sources:}    
    real_source = quantile(ref_im_warp$imDat, source_threshold, na.rm = TRUE)
\textcolor{cyan}{#Note, na.rm=TRUE means `NA' flagged pixels are ignored} 
\textcolor{cyan}{#Step 2ii: Find real sources}
    relflux = relflux[ref_im_warp$imDat > real_source]
    
\textcolor{cyan}{#Step 2iii: Only consider positive relative fluxes}
    relflux = relflux[relflux > 0] #ignore negative values
    
\textcolor{cyan}{#Step 2iv: Define very blue (but real) pixels}
    scale = quantile(relflux, scale_threshold, na.rm = TRUE)
  
\textcolor{cyan}{#Step 3: Subtract scaled reference image}
    wisp_template = wisp_im$imDat - ref_im_warp$imDat*scale

\textcolor{cyan}{#Step 4: Clip extreme pixels}
wisp_template[wisp_template > quantile(wisp_template[ \textcolor{blue}{\it cont...}
        wisp_template > 0], clip_threshold, na.rm = TRUE)] = NA
        
\textcolor{cyan}{#Step 5: Smooth the template}
    wisp_template = profoundImBlur(wisp_template, sigma=sigma)
    
\textcolor{cyan}{#Step 6: Set to NA pixels we do not want to consider}
    wisp_template[is.na(wisp_im$imDat) | is.na(ref_im_warp$imDat)] = NA
    wisp_template[wisp_template < 0] = NA

\textcolor{cyan}{#Step 7: Smooth the template again}
    wisp_template_lo = profoundImBlur(wisp_template, sigma = sigma_lo)
    
\textcolor{cyan}{#Step 8: Combine high and low frequency templates}
    sel = which(is.na(wisp_template) | wisp_template_lo > wisp_template)
    wisp_template[sel] = wisp_template_lo[sel]
 
\textcolor{cyan}{#Step 9: Subtract wisp model from the image we wish to fix}
    wisp_im$imDat  = wisp_im$imDat - wisp_template
  
\textcolor{cyan}{#Return the wisp corrected image}
    return(wisp_im)
\}
\end{alltt}

\end{document}